\documentclass[3p,times,procedia]{elsarticle}
\usepackage{nupha_ecrc}

\volume{00}

\firstpage{1}

\journalname{Nuclear Physics A}

\runauth{}

\jid{nupha}

\jnltitlelogo{Nuclear Physics A}

\usepackage{amssymb}

\usepackage[figuresright]{rotating}

\newcommand{\rp}{{\rm RP}}

\begin{document}

\begin{frontmatter}



\dochead{XXVIth International Conference on Ultrarelativistic Nucleus-Nucleus Collisions\\ (Quark Matter 2017)}

\title{Chiral magnetic effect in isobaric collisions}


\author[fudan1,fudan2]{Xu-Guang Huang}
\author[hust]{Wei-Tian Deng}
\author[sinap]{Guo-Liang Ma}
\author[ucla]{Gang Wang}

\address[fudan1]{Physics Department and Center for Particle Physics and Field Theory, Fudan University, Shanghai 200433, China}
\address[fudan2]{The Key Lab of Applied Ion Beam Physics, Ministry of Education, Fudan University, Shanghai 200433, China}
\address[hust]{School of physics, Huazhong University of Science and Technology, Wuhan 430074, China}
\address[sinap]{Shanghai Institute of Applied Physics, Chinese Academy of Sciences, Shanghai 201800, China}
\address[ucla]{Department of Physics and Astronomy, University of California, Los Angeles, California 90095, USA}

\begin{abstract}
We give a numerical simulation of the generation of the magnetic field and the charge-separation signal due to the chiral magnetic effect (CME) --- the induction of an electric current by the magnetic field in a parity-odd matter --- in the collisions of isobaric nuclei, $^{96}_{44}$Ru + $^{96}_{44}$Ru and $^{96}_{40}$Zr + $^{96}_{40}$Zr, at $\sqrt{s_{\rm NN}}=200$ GeV. We show that such collisions provide an ideal tool to disentangle the CME signal from the possible elliptic-flow driven background effects. We also discuss some other effects that can be tested by using the isobaric collisions.
\end{abstract}

\begin{keyword}
Chiral magnetic effect \sep isobaric collisions
\end{keyword}

\end{frontmatter}


\section{Introduction}
\label{sec:intro}
The chiral magnetic effect (CME) represents the generation of an electric current induced by the magnetic field in a parity-odd environment~\cite{Kharzeev:2007jp,Fukushima:2008xe}. In the hot quark-gluon matter generated in high-energy heavy-ion collisions, such a parity-odd environment may be produced from the vacuum transition induced by topologically nontrivial gluon fields, e.g., sphalerons. Thus the observation of the CME in heavy-ion collisions could provide a means to monitoring the topological sector of quantum chromodynamics (QCD). In recent years, the experimental search for the CME has been intensively performed in heavy-ion collisions at the RHIC and the LHC and encouraging results consistent with the expectation of CME were indeed observed. However, it is known that several elliptic-flow driven effects which are independent of the topological transition in QCD could possibly lead to similar results and therefore make the interpretation of the experimental data ambiguous.

Let us first briefly discuss the experimental observable to detect the CME and its background contributions; more information can be found in, e.g., Refs.~\cite{Huang:2015oca,Hattori:2016emy,Kharzeev:2015znc,Wang:2016mkm}. In the experiments of heavy-ion collisions, a three-point correlator,
\begin{eqnarray}
\label{corre}
\gamma_{\alpha\beta}=\langle\cos(\phi_\alpha+\phi_\beta-2\Psi_\rp)\rangle,
\end{eqnarray}
was designed to detect the CME~\cite{Voloshin:2004vk}, where $\phi_\alpha, \phi_\beta$ ($\alpha,\beta=\pm$ is charge sign) are the azimuthal angles of the charged particles, $\Psi_\rp$ is the angle of the reaction plane of a given event, and $\langle\cdots\rangle$ denotes an average over all particle pairs and all the events. The CME would drive a charge separation with respect to the reaction plane and thus contribute a positive opposite-sign (OS) correlator and a negative same-sign (SS) correlator --- a pattern indeed observed by the STAR Collaboration for Au + Au collisions as well as for Cu + Cu or U + U collisions~\cite{Abelev:2009ac,Abelev:2009ad,Wang:2012qs,Tribedy,Adamczyk:2014mzf} and
by the ALICE Collaboration for Pb + Pb collisions at $\sqrt{s_{\rm NN}}=2.76$ TeV~\cite{Abelev:2012pa}. However, there exist ambiguities in the interpretation of the experimental results, as possible background effects
that are not related to the CME, once coupled with elliptic flow ($v_2$), could also contribute to $\gamma$. Such effects include, e.g., local charge conservation~\cite{Pratt:2010gy,Schlichting:2010qia}, neutral resonance decays~\cite{Wang:2009kd}, and transverse momentum conservation~\cite{Pratt:2010gy,Pratt:2010zn,Bzdak:2010fd}.

One way to disentangle the possible CME signal and the flow-related backgrounds is to use the prolate shape of the uranium nuclei~\cite{Voloshin:2010ut}: In central U + U collisions, one expects sizable $v_2$ but a negligible magnetic field,
and thus a vanishingly small CME contribution to the correlator $\gamma$. The corresponding measurement was performed by the STAR Collaboration in 2012 and they indeed found sizable $v_2$ while
the difference between $\gamma_{\rm OS}$ and $\gamma_{\rm SS}$, $\Delta\gamma\equiv\gamma_{\rm OS}-\gamma_{\rm SS}$,
is consistent with zero~\cite{Wang:2012qs,Tribedy}. More discussion can be found in Refs.~\cite{Wang:2012qs,Tribedy}.
Another way is to vary the magnetic field with the backgrounds fixed~\cite{Voloshin:2010ut}. This comes the idea of isobaric collisions.

\section{Isobaric collisions and CME}
\label{sec:iso}
Isobaric nuclei have the same atomic number but different charge numbers, e.g., the nuclei of $^{96}_{44}$Ru and $^{96}_{40}$Zr both have $96$ nucleons but the former contains $10\%$ excess charges than the latter. Thus, at given beam energy and centrality, $^{96}_{44}$Ru + $^{96}_{44}$Ru collisions would generate roughly the same $v_2$ but $10\%$ larger magnetic field than $^{96}_{40}$Zr + $^{96}_{40}$Zr collisions. Therefore, one expects $\Delta\gamma$ in Ru + Ru collisions to be roughly $20\%$ larger than that in Zr + Zr collisions {\it if} $\Delta\gamma$ is dominated by CME. On the other hand, if $\Delta\gamma$ is dominated by background effects, it will not show significant difference in the two collisions. In the following, we give our detailed numerical study following Ref.~\cite{Deng:2016knn}.

We model the nucleon distribution of either Ru or Zr by the Woods-Saxon form (in rest frame),
\begin{eqnarray}
\rho(r,\theta)=\frac{\rho_0}{1+\exp{[(r-R_0-\beta_2 R_0 Y^0_2(\theta))/a]}},
\end{eqnarray}
where $\rho_0=0.16$ fm$^{-3}$, $R_0$ and $a$ are the ``radius" and the surface diffuseness parameter, respectively, and $\beta_2$ is the deformity of the nucleus. The parameter $a$ is almost identical for Ru and Zr: $a\approx 0.46$ fm. The current information for $\beta_2$ is ambiguous~\cite{Shou:2014eya}: The e-A scattering experiments~\cite{Raman:1201zz,Pritychenko:2013gwa} give $\beta_2^{\rm Ru} = 0.158$ and $\beta_2^{\rm Zr} = 0.08$ (which will be referred to as case 1) while the comprehensive model deductions~\cite{Moller:1993ed} give $\beta_2^{\rm Ru}=0.053$ and $\beta_2^{\rm Zr}=0.217$ (which will be referred to as case 2).

\begin{figure}[!htb]
\begin{center}
\includegraphics[width=6.5cm]{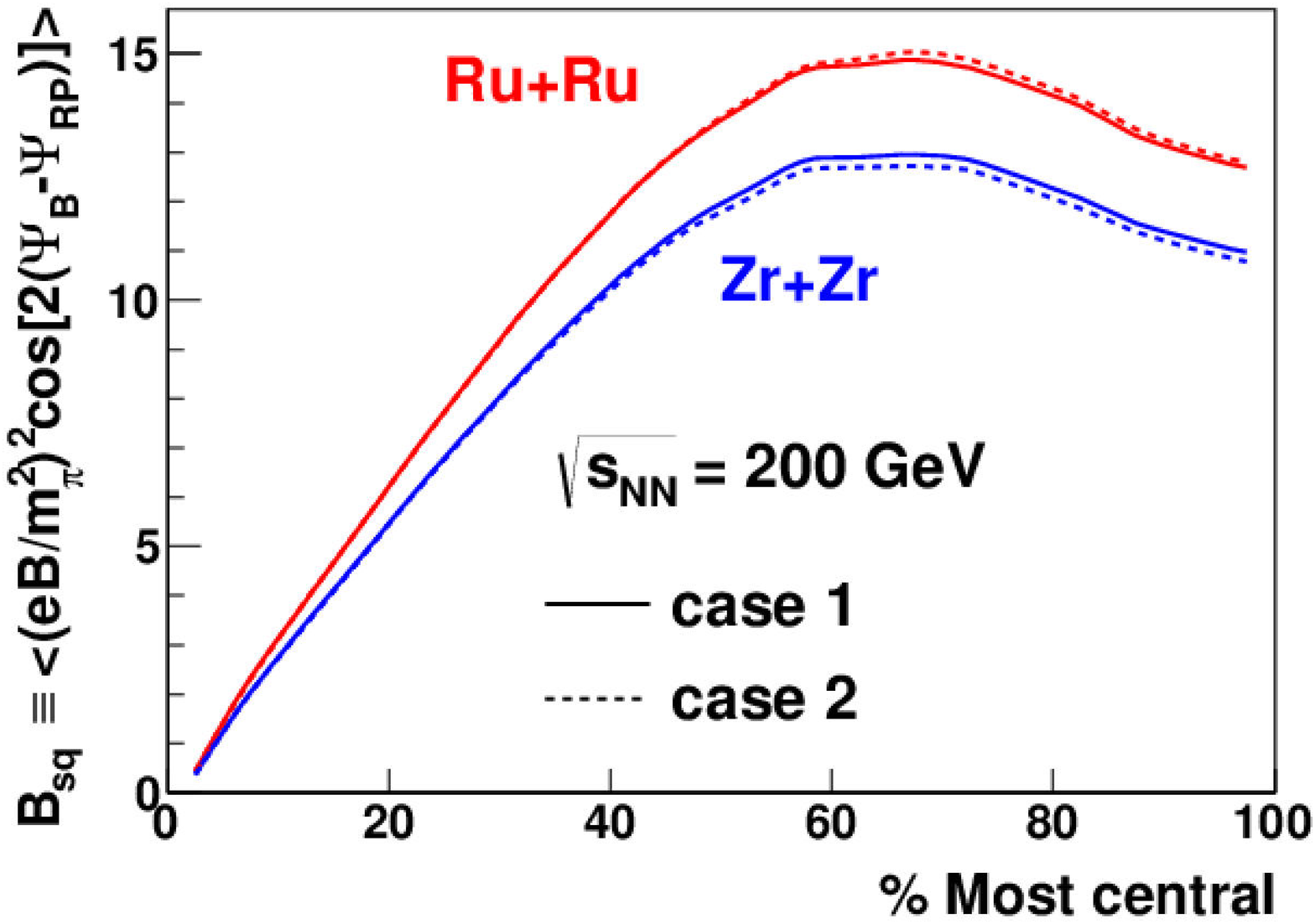}
\includegraphics[width=6.5cm]{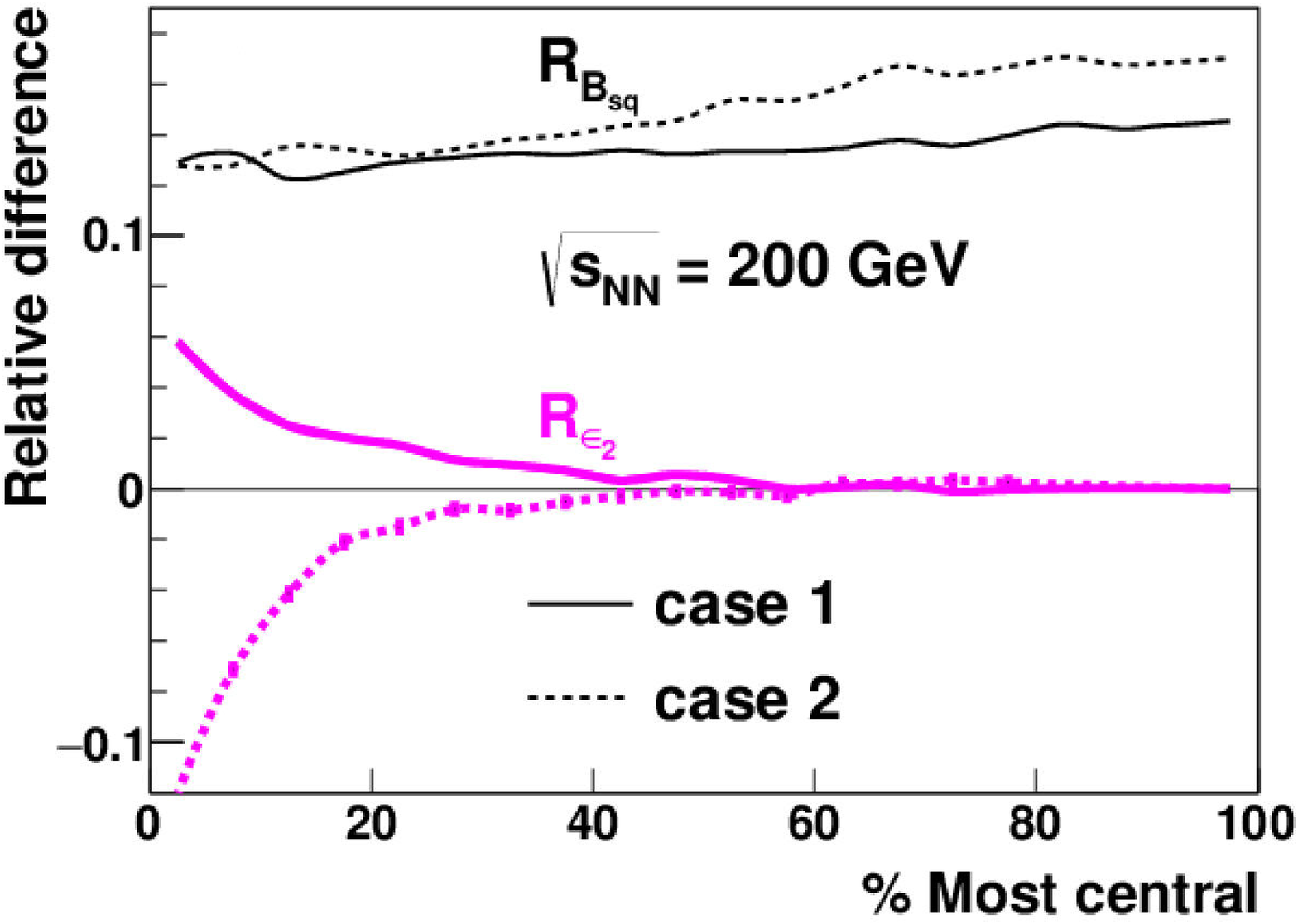}
\caption{The event-averaged projected magnetic field squared at the center of mass of the overlapping region at $\sqrt{s_{\rm NN}}=200$ GeV (Left) and their relative difference (Right) versus centrality. The pink curves are the relative difference in initial eccentricity (Right).}
\label{fig_mag}
\end{center}
\end{figure}
We in Fig.~\ref{fig_mag} (Left) show the numerical result for the event-averaged projected initial magnetic field squared, $B_{sq}\equiv\langle(eB/m_\pi^2)^2\cos[2(\Psi_{\rm B}-\Psi_{\rm RP})]\rangle$ (with $m_\pi$ the pion mass and $\Psi_{\rm B}$ the azimuthal angle of the magnetic field), at the center of mass of the overlapping region for the two collision systems at 200 GeV, using the HIJING model~\cite{Deng:2012pc,Deng:2014uja}. This quantity characterizes the magnetic field's capability of driving the CME signal in $\gamma$~\cite{Bloczynski:2012en,Bloczynski:2013mca}. Obviously, for the same centrality, the $B_{sq}$ in Ru + Ru collisions is bigger than in Zr + Zr collisions. The relative difference in $B_{sq}$, $R_{B_{sq}} \equiv 2(B_{sq}^{\rm Ru+Ru}-B_{sq}^{\rm Zr+Zr})/(B_{sq}^{\rm Ru+Ru}+B_{sq}^{\rm Zr+Zr})$ (similarly for $R_{\epsilon_2}$, $R_{S}$ etc, below), approaches $15\%$ (case 1) or $18\%$ (case 2) for peripheral events, and reduces to about $13\%$ (case 1 and case 2) for central events, as seen in Fig.~\ref{fig_mag} (Right). On the other hand, the relative difference in the initial eccentricity, $R_{\epsilon_2}$, obtained from the
Monte Carlo Glauber simulation, is always much smaller than $R_{B_{sq}}$, as shown in the pink curves in Fig.~\ref{fig_mag} (Right). This indicates that the $v_2$-driven effects should stay almost the same (particularly for centrality bins $>20\%$) while the magnetic-field induced effect should behave quite differently between Ru + Ru and Zr + Zr collisions.

\begin{figure}[!htb]
\begin{center}
\includegraphics[width=6.5cm]{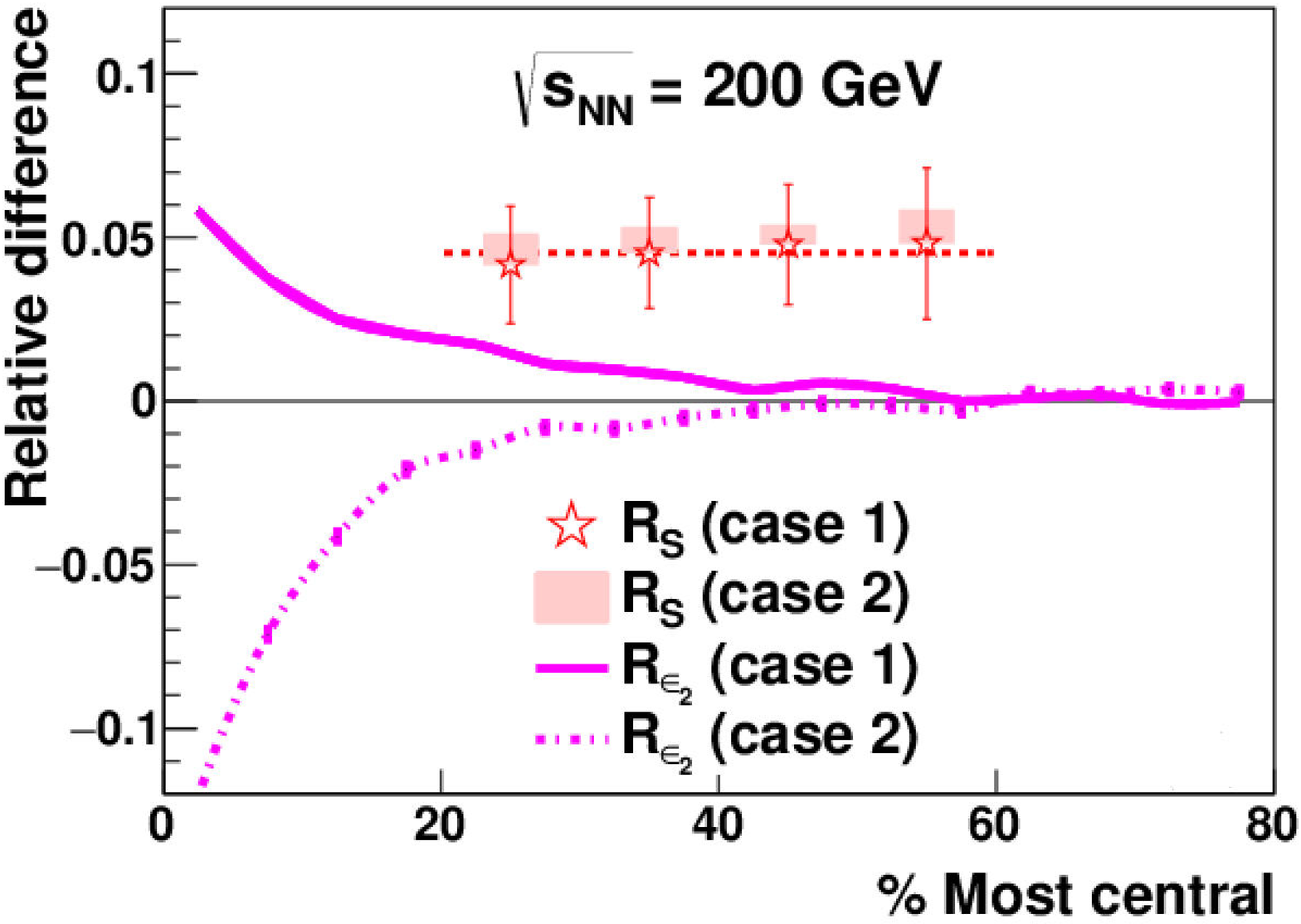}
\includegraphics[width=7.5cm]{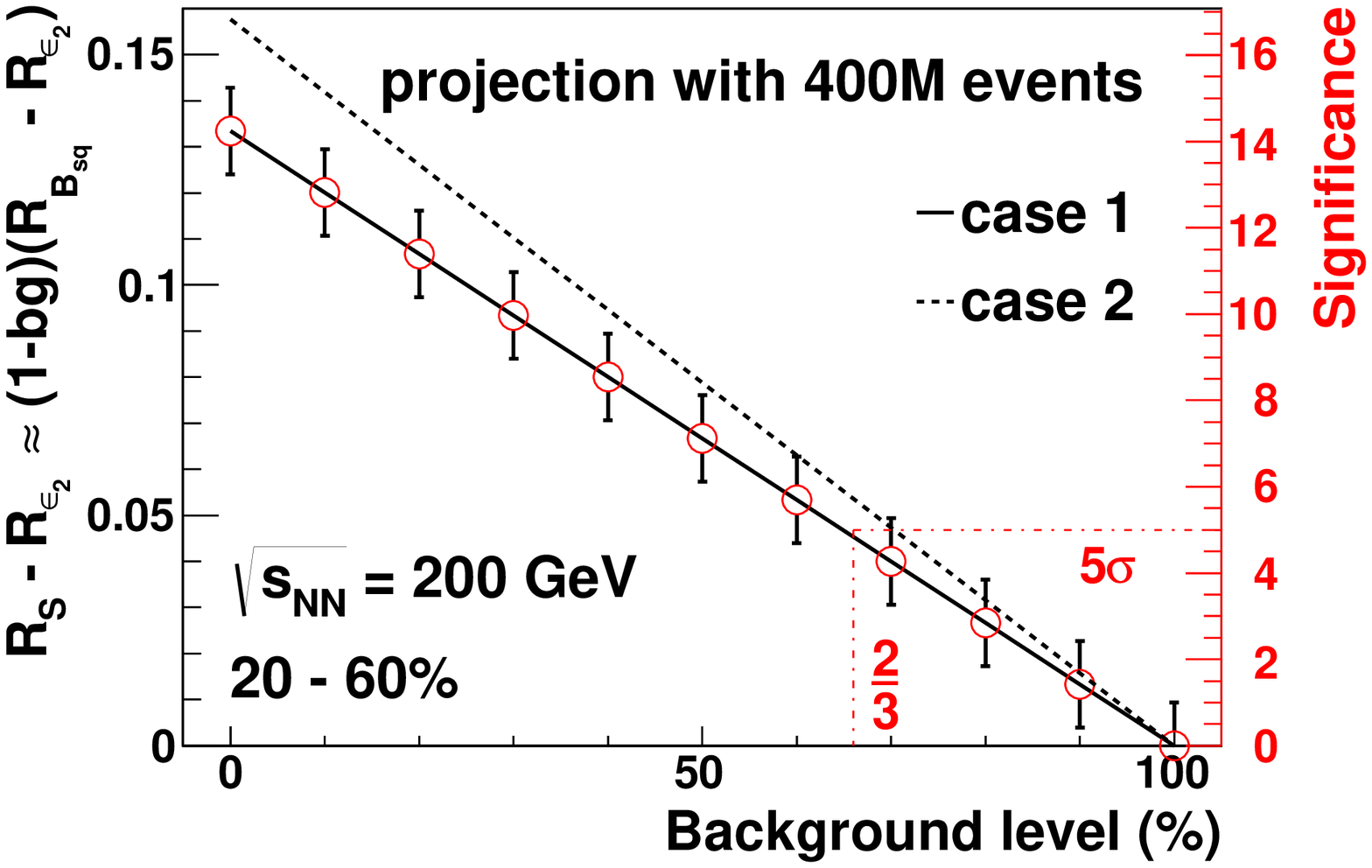}
\caption{The relative difference in $S$ between Ru + Ru and Zr + Zr collisions at $200$ GeV in the centrality range $20-60\%$ (Left). Magnitude (left axis) and significance (right axis) of the relative difference, $R_S-R_{\epsilon_2}$, in the CME signal at $200$ GeV, as a function of the background level.}
\label{fig_res}
\end{center}
\end{figure}

Given the initial magnetic fields and eccentricities, we now turn to discuss the charge-separation observable
$S\equiv N_{\rm part}\Delta\gamma$, where $N_{\rm part}$ is used to compensate for the possible dilution
effect~\cite{Abelev:2009ad,Ma:2011uma}. For this purpose, we take a two-component perturbative approach to the relative difference in $S$~\cite{Bzdak:2012ia},
\begin{eqnarray}
R_S = (1-{\rm bg})R_{B_{sq}}+ {\rm bg}\cdot R_{\epsilon_2},
\end{eqnarray}
where we introduce ${\rm bg}\in [0,1]$ to describe the background level. In Fig.~\ref{fig_res} (Left) we show $R_S$ for the centrality range of $20-60\%$ with ${\rm bg}=2/3$ and under the statistics of $400\times 10^{6}$ events; in comparison, we show $R_{\epsilon_2}$ again. For both case 1 (red stars) and case 2 (pink shaded boxes) the relative difference in $S$ is about $5\%$ in the plotted centrality range. We checked that when we combine the events of $20-60\%$ centralities, $R_S$ is $5\sigma$ above $R_{\epsilon_2}$ for both cases. We therefore conclude that the isobaric collisions provide a very promising test to pin down the underlying mechanism for the observed charge separation. In Fig.~\ref{fig_res} (Right) we plot the relative difference in the CME signal, i.e., $R_S-R_{\epsilon_2}$, at 200 GeV with the statistics of $400\times 10^{6}$ events, as a function of the background level ${\rm bg}$. Such a plot will be useful for determining the background level when compared with the future experimental results.

\section{Testing other effects with isobaric collisions}
\label{sec:by}
Although the primary aim of running the isobaric collisions is to test the CME, there are other interesting effects that can be tested by the isobaric collisions. Here we list two of them. \\
(1) As we mentioned in last section, the current knowledge of the deformity (reflected in parameter $\beta_2$) of Ru and Zr nuclei is ambiguous: In case 1, Ru nucleus is more deformed while in case 2, the Zr nucleus is more deformed. As shown in Fig.~\ref{fig_mag} (Right), the two cases lead to opposite trends in $R_{\epsilon_2}$ and thus would lead to opposite trends in $R_{v_2}$ in the central events. Thus the $v_2$ measurements in central isobaric collisions will be useful to discern which nucleus is more deformed. \\
(2) Recently, the global spin polarization of $\Lambda$ or $\bar\Lambda$ baryon was measured by the STAR Collaboration in Au + Au collisions~\cite{STAR:2017ckg,Upsal}. The result shows evident splitting between the polarizations of $\Lambda$ and $\bar\Lambda$ at energies lower than 39 GeV. One possible mechanism for this splitting is the magnetic field: $\Lambda$ and $\bar\Lambda$ have opposite magnetic moments and thus the magnetic field can lead to opposite spin polarizations to them. Running isobaric collisions at lower energies will provide a rare opportunity to test whether the magnetic field is the dominant contribution to the spin polarization splitting between $\Lambda$ and $\bar\Lambda$.

\section{Summary}
\label{sec:summ}
We calculated numerically the initial magnetic field and its event-by-event fluctuation (reflected in $B_{sq}$) and the initial eccentricity of the overlapping region for isobaric collisions, Ru + Ru and Zr + Zr. We show that the isobaric collisions can provide a valuable opportunity to disentangle the CME signal from the elliptic-flow driven effects. The isobaric collisions may also be used to test some other interesting phenomena, like determining the deformity of Ru and Zr, or testing the spin polarization splitting between $\Lambda$ and $\bar\Lambda$.

{\bf Acknowledgments:} This work is supported by NSFC with Grant No. 11405066 (W.-T.D), No. 11535012 and No. 11675041 (X.-G.H), No. 11375251, No. 11522547, and No. 11421505 (G.-L.M), and the US Department of Energy under Grant No. DE-FG02-88ER40424 (G.W).


\end{document}